\documentclass[11pt,twoside]{article}

\usepackage{asp2006}
\usepackage{graphicx}
\usepackage{psfig}

\newcommand\simlt{\lower.5ex\hbox{$\; \buildrel < \over \sim \;$}}
\newcommand\simgt{\lower.5ex\hbox{$\; \buildrel > \over \sim \;$}}

\begin{document}
\title{Gamma Ray and Neutrino Emission as a Probe of Relativistic Jets}   
\author{Amir Levinson}   
\affil{School of Physics and Astronomy, Tel Aviv University, Tel Aviv 69978, Israel}    

\begin{abstract} 
Constraints on the dynamics, dissipation, and production of VHE neutrinos in relativistic jets are derived using 
opacity calculations and VHE $\gamma$-ray observations.  In particular, it is demonstrated how rapid 
variability of the $\gamma$-ray emission at very high energies ($> 100$ GeV) can be used to map the location of the 
$\gamma$-spheres, to derive lower limits on the Doppler factor of the $\gamma$-ray emission zone, and to
constrain the photopion production opacity.    The apparent discrepancy between jet Lorentz factors inferred from 
 superluminal motions and source statistics in the TeV blazars and those derived from the $\gamma$-ray emission is discussed. 
The relation to the high-energy emission from the HST1 knot in M87 is briefly mentioned. 
Estimates of neutrino yields in upcoming neutrino telescopes are given
for various sources.  It is shown that for TeV blazars, the rapid variability of the TeV emission implies neutrino yields 
well below detection limit.

\end{abstract}


\section{Introduction}   
There is now strong evidence that the high-energy emission observed in 
blazars, microquasars, and $\gamma$-ray bursts originates from collimated, 
relativistic outflows. The common view is that those outflows are powered 
by a magnetized accretion disk and a spinning black hole and collimated by 
magnetic fields and/or the medium surrounding the jet.  However, there is 
as yet no universal agreement about the mechanisms responsible for the formation and 
acceleration the jet and the dissipation of its bulk energy.  Even the composition of the jet is unknown in most sources.  The rapid variability frequently observed suggests that the high-energy emission 
is likely to be produced close to the inner engine and, therefore, provides an important probe of the 
physics involved on those scales.  As discussed below, some important constraints on the jet parameters can already be imposed using available data, and further progress is expected in the coming few years with the advance of observational techniques.  A new generation of experiments just started operating or will become operative soon: Space-based (e.g. GLAST, INTEGRAL, AGILE), Cerenkov
(e.g. HESS, MAGIC, VERITAS), and air-shower (e.g. MILAGRO)
$\gamma$-ray detectors will probe with high sensitivity the energy
range of 10~MeV to a few TeV (see, e.g., Catanese, 1999 for a review). 
The operating BAIKAL and AMANDA neutrino telescopes,
and the cubic-km scale telescopes under construction (IceCube,
ANTARES, NESTOR, NEMO; see, e.g. Halzen, 2005 for review),
will open a new window onto the Universe. Besides providing an important probe of the 
innermost regions of compact astrophysical systems, these experiments can also be exploited to 
test new physics.   Finally, new ultra-high-energy cosmic-ray detectors, e.g. the 
HiRes and the hybrid Auger detectors will provide cosmic-ray data of unprecedented quality and quantity. 

\section{Gamma Rays}

At sufficiently small radii the {\em compactness parameter}, 
$l_\gamma=L_\gamma\sigma_T/(m_ec^2r)\simeq10^4(L_\gamma/L_{\rm Edd})(r/r_s)^{-1}$, largely exceeds
unity for essentially all classes of compact high-energy sources.  It is, therefore, expected that
$\gamma$-rays will not be able to escape from the inner jet regions without creating pairs.
Both the synchrotron photons produced inside the jet and the ambient radiation intercepted by the 
jet contribute an opacity to pair production.  The threshold $\gamma$-ray energy for pair production 
via interaction with a target photon of energy $h\nu_s$ is
\begin{equation}
\epsilon_{\gamma,th}\simeq 2.5\times10^{11}
\left({h\nu_s\over {\rm 1 eV}}\right)^{-1}\qquad {\rm eV}.
\label{thrs}
\end{equation}
Consequently, for a target photon spectrum $F_\nu\propto\nu_s^{-\alpha}$ with $\alpha>0$ the $\gamma$-ray opacity at a given radius should increase with increasing $\gamma$-ray energy.  Results of detailed calculations of the $\gamma$-ray opacity are exhibited in figure 1, where the {\em $\gamma$-spheric radius}, defined as the radius 
$r_{\gamma}(\epsilon_{\gamma})$ beyond which the pair production optical depth to infinity
is unity, viz., $\tau_{\gamma\gamma}(r_\gamma,\epsilon_\gamma)=1$, is 
plotted against $\gamma$-ray energy $\epsilon_{\gamma}$, for two target radiation fields: 
synchrotron radiation (dashed lines) and external radiation (solid lines).  The spectra of the target
radiation fields employed in those calculations are discussed in Levinson (2006).  As seen
the $\gamma$-spheric radius indeed increases with increasing $\gamma$-ray energy.

The $\gamma$-spheres can be mapped in principle by measuring
temporal variations of the $\gamma$-ray flux in different energy bands during a flare.
If the $\gamma$-ray emission is produced over many octaves of jet radius, where 
intense pair cascades at the observed energies are important (Blandford \& Levinson 1995), then 
it is expected that a flare will propagate from low to high energies, or that the 
variations at higher $\gamma$-ray energy will be slower than at lower energies.
With the limited sensitivity and energy band of the EGRET instrument it was practically 
impossible to resolve such effects.  It is hoped that with the upcoming GLAST instrument this will be feasible

\begin{figure}[h]
\includegraphics[width=13.0cm]{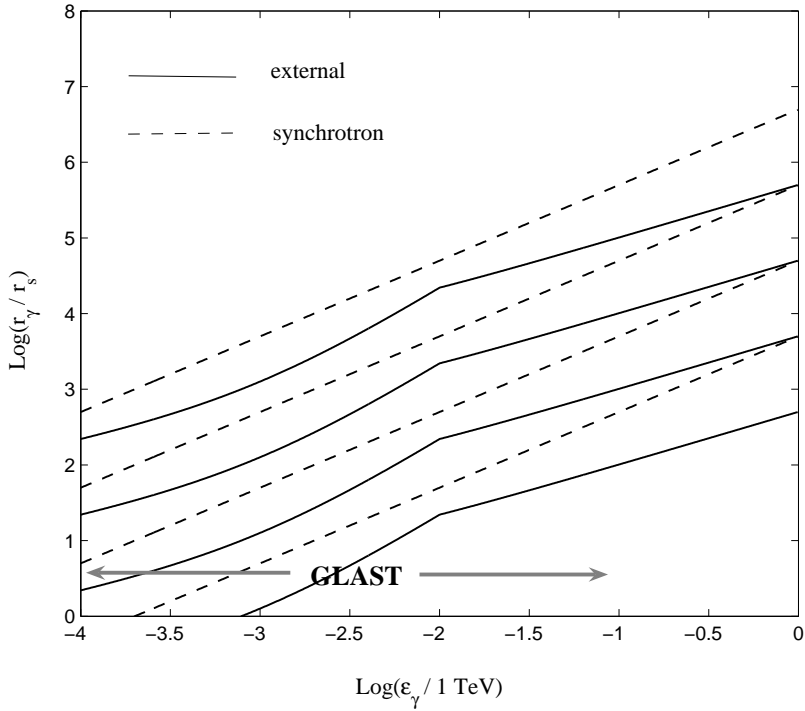}
\caption{Dimensionless $\gamma$-spheric radius versus $\gamma$-ray energy, computed using the target photon spectra 
from Levinson (2006).  The different curves correspond to a different normalization of the target radiation field intensity.}
\end{figure}
 
Further constraints on the source parameters can be derived by measuring the low-energy flux 
(radio-to-IR) simultaneously with the variable $\gamma$-ray emission.  Suppose that a variability timescale $\Delta t$
has been measured at some observed $\gamma$-ray energy $\epsilon_\gamma$.
This implies  that the emission at the observed energy originated from a region of size (as measured in the Lab frame)
$\Delta r< r_{\rm var}\equiv\Gamma\delta c \Delta t/(1+z)$, where $\Gamma$ and $\delta$ are the bulk Lorentz factor 
and the corresponding Doppler factor of the emitting matter, respectively, and $z$ is the redshift of the source. 
Above the emission region the pair production optical depth at the observed energy 
must not exceed unity, viz., $\tau^{\rm syn}_{\gamma\gamma}(r_{\rm em},\epsilon_\gamma/\delta)<1$,
where $r_{\rm em}$ is the radius at which the $\gamma$-ray emitting region is located. 
Assuming $r_{\rm em}\simeq\Delta r<r_{\rm var}$ implies $r_{\gamma}(\epsilon_\gamma)< r_{\rm var}$.  Suppose now that the 
synchrotron flux near the peak can be measured simultaneously.  
The latter condition on $r_{\gamma}(\epsilon_\gamma)$ can be solved for the Doppler factor to yield (Levinson 2006)
\begin{equation}
\delta^{5}>2\times10^{11}\Delta 
t_h^{-3/2}(\Gamma\theta)^{-2}z^2(\epsilon_\gamma/{\rm 1 TeV})^{1/2}S_{{\rm Jy}},
\label{Var-const}
\end{equation}
where $S_{\rm Jy}$ is the measured synchrotron flux density in Janskys,
$\theta$ is the opening angle of the jet and $\Delta t_h$ is $\Delta t$ measured in hours.  Adopting $\Gamma\theta\simeq1$, we estimate $\delta>35$ for the rapid flare observed in Mrk 421 and $\delta>190$ for the few minuets variability reported for PKS 2155-304, with $r_{\rm em}<10^{17}$ cm for the minimum condition in both sources.
Such high values of $\delta$ are in clear disagreement with the much lower values inferred from 
source statistics (e.g., Urry \& Padovani 1991)  and superluminal motions on parsec scales (Jorstad et al. 2001).
Various explanations, including jet deceleration (Georganopouloselta \& Kazanas 2003), a structure 
consisting of interacting spine and sheath (Ghisellini et al. 2005), and opening angle 
effects (Gopal-Krishna et al. 2004) have been proposed in order to resolve this discrepancy.  Alternatively, it is conceivable that those flares result from impulsive ejection episodes of high Lorentz factor shells, like in the internal shock model of GRBs.

It has been argued that such high values of the Doppler factor may not be required if the 
$\gamma$-ray production region is located far from the black hole, at radii $r_{\rm em}>>\Delta r$.
However, the fraction of jet energy that can be tapped for production of $\gamma$ rays in a region of 
size $\Delta r$ located at a radius $r_{\rm em}$ is $\eta\sim (\Delta r/\theta r_{em})^2$.  As a consequence, 
either the opening angle of the jet is very small, $\theta \sim\Delta r/r<<1$ or the jet power is much larger than the luminosity of the TeV emission measured during the flaring states.  Another possibility is that the TeV emission is produced by a converging shock in a reconfinement nozzle, as proposed for the HST1 knot in M87 (Cheung et al. 2007).  It should be noted that in M87 the X-ray and TeV luminosities 
are $L_{\rm TeV}\sim L_{\rm x}\simlt10^{41}$ erg s$^{-1}$ (Aharonian et al. 2006; Cheung et al. 2007), much smaller than the TeV luminosity $L_{\rm TeV}\sim 10^{44-45}$ erg s$^{-1}$ observed typically in the TeV blazars.  Estimates of the jet power in M87 
yield $L_j\simgt10^{44}$  erg s$^{-1}$ (e.g., Stawarz et al. 2006; Bicknell \& Begelman 1996), implying a very small conversion fraction, $L_{\rm TeV}/L_j\simlt10^{-3}$.  Even with such a small conversion efficiency an opening angle $\theta<10^{-2}$ rad is required 
if the TeV emission were to originate from the HST1 knot, unless reconfinement, as mentioned above, can give rise to convergence of the jet.  This may occur if radiative cooling of the shocked jet layer is effective.  Recollimation shocks may be an important dissipation channel also in other sources, e.g., GRBs (Bromberg \& Levinson, 2007)

\section{Cosmic rays and Neutrinos}
A substantial fraction of the energy dissipated in the jet can, in principle, be tapped for the acceleration 
of protons to ultrahigh energies with a power law spectrum.
The maximum proton energy is limited by energy losses at small radii and by escape at larger radii (Levinson 2006), and 
in any case is restricted to $\epsilon_{p,max}\le eB\theta r$, where $\theta$ is the opening angle of the jet and 
$B$ the magnetic field.  In terms of the equipartition parameter $\xi_B$ the magnetic field can be expressed 
as $B=4\times10^8(\xi_B{\cal L}_j/m_{BH})^{1/2}(\theta r)^{-1}$ Gauss, where ${\cal L}_j$ is the jet power in Eddington units and $m_{BH}$ is the mass of the black hole in units of solar mass. Consequently, production of UHECRs 
of observed energy $\epsilon_p$ requires
\begin{equation}
\xi_B m_{BH}{\cal L}_j\ge10^8(\epsilon_p/10^{20.5}{\rm eV})^2,
\label{Hillas}
\end{equation} 
which essentially leaves GRBs with $m_{BH}\sim3$, $\xi_B\simeq0.1$, ${\cal L}_j\simeq 10^{12}$,
and powerful blazars with $m_{BH}\sim10^9$, $\xi_B\simeq0.1$, ${\cal L}_j\simeq 1$
as the main candidates for astrophysical UHECRs sources\footnote{
Condition (\ref{Hillas}) does not apply to vacuum gaps in dormant AGNs (Boldt \& Gosh 1999; Levinson 2000) and magnetars
where the ideal MHD condition is violated.}  Whether protons can actually be accelerated by 
relativistic shocks to the highest energies observed is yet an open issue.  

High-energy neutrinos can be produced in astrophysical jets mainly through the 
decay of charged pions.   The pions may be produced through collisions of protons with 
target photons ($p\gamma$), or via inelastic ($pp$) and ($pn$) collisions.  The former process
is usually dominant in jets (for some exceptions see, Levinson \& Eichler, 2003; Torres \& Halzen, 2007).

Observations of VHE $\gamma$-rays can constrain the photopion production 
opacity, particularly in situations where rapid variability of 
the VHE $\gamma$-ray flux is observed.  Because both the protons and the $\gamma$ rays interact 
locally with the same target radiation field the ratio of photomeson and pair production opacities 
depend solely on the ratio of cross sections, $\sigma_{p\gamma}/\sigma_{\gamma\gamma}\simeq 4\times 10^{-3}$,
and the spectrum of the target radiation field.  For a target photon spectrum $n_s(\nu)\propto \nu^{-\alpha}$
this ratio reads
\begin{equation}
{\tau_{p\gamma}(\epsilon_p,r)\over \tau_{\gamma\gamma}(\epsilon_{\gamma},r)}
=\frac{n_s(\epsilon_{p,th})}{n_s(\epsilon_{\gamma,th})}
\frac{\sigma_{p\gamma}}{\sigma_{\gamma\gamma}}\simeq4\times10^{-3}
\left(\frac{\epsilon_p}{3\times10^5\epsilon_{\gamma}}\right)^{\alpha}.
\label{tau_pg}
\end{equation}
Detailed calculations of opacity ratios that employed more realistic target photon spectra
are presented in Levinson (2006), and are plotted in fig. 2.  As seen from the figure, at $\gamma$-ray energies above 
a few TeV the opacity ratio is smaller than unity even at the maximum proton energy.  For the class of TeV blazars 
this implies neutrino yields well below detection limit (see Levinson 2006 for more details).
With GLAST it should be possible to constrain other sources and to use such constraints to identify the best candidates
for the upcoming km$^3$ detectors.
\begin{figure}[h]
\includegraphics[width=14.0cm]{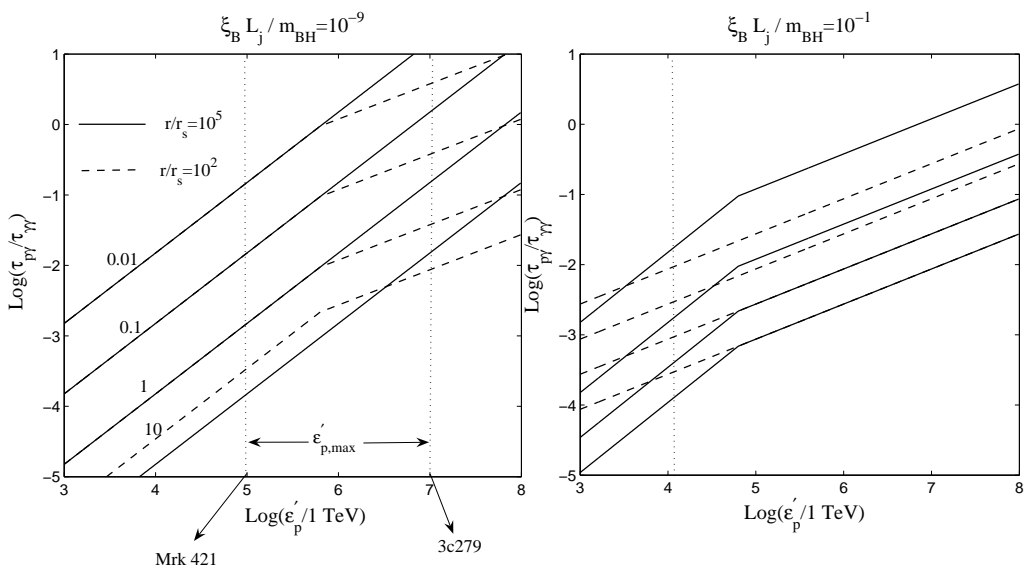}
\caption{Opacity ratio, $\tau^{syn}_{p\gamma}(\epsilon_p^\prime)/
\tau^{syn}_{\gamma\gamma}(\epsilon_{\gamma}^\prime)$,
as a function of comoving proton energy $\epsilon_p{^\prime}$ at two radii: $r/r_s=10^2$ (dashed lines)
and $10^5$ (solid lines).  The left panel corresponds to a choice of parameters that represents a 
typical blazar like 3c279, and the right panel to a choice of parameters representing a typical microquasar like GRS 1915.  The numbers that label the curves correspond to the comoving $\gamma$-ray
energy in units of TeV.  The dotted lines designate the maximum proton energy derived from the confinement limit.}
\end{figure}

The neutrino flux emitted from the jet depends on several factors:
(i) the fraction of jet energy injected as a power law distribution of 
protons $\xi_p$, (ii) the spectrum of accelerated protons, (iii) 
the cooling times of pions and muons and (iv) the pion production opacity.
It is customary to introduce the parameter $f_\pi(\epsilon_p^\prime)\le1$, 
representing the fraction of proton energy $\epsilon_p^\prime$ 
converted to pions.    In case of photopion production it can be approximated
by $f_\pi(\epsilon_p^\prime)={\rm min}[1,K_\pi\tau_{p\gamma}(\epsilon_p^\prime)]$, where $K_\pi$ is the elasticity.
For a proton distribution $n_p(\epsilon_p^\prime)$ the average fraction of proton energy
lost to pions can be defined as 
\begin{equation}
\bar{f}_\pi={\int f_\pi(\epsilon_p^\prime)\epsilon_p^\prime n_p(\epsilon_p^\prime)
d\epsilon_p^\prime\over \int \epsilon_p^\prime n_p(\epsilon_p^\prime)
d\epsilon_p^\prime}.
\label{f_pi}
\end{equation}
This average fraction depends on the spectrum of both protons and target photons.
For powerful blazars this fraction is estimated to be $\bar{f}_\pi\simgt0.2$ assuming a proton 
energy distribution $n_p(\epsilon)\propto \epsilon_p^{-2} $, whereas for the TeV BL Lac sources
$\bar{f}_\pi<10^{-4}$ (Levinson 2006).  The corresponding event rate in a km$^3$ detector is 
$\dot{N}_\mu\sim10\xi_p$ yr$^{-1}$ for a powerful blazar at a redshift $z=1$, and $\dot{N}_\mu\simlt 0.03\xi_p$
yr$^{-1}$ for Mrk 421 and Mrk 501.
In the case of microquasars up to a few events can be detected during a strong outburst if the viewing angle is sufficiently small (Levinson \& Waxman 2001; Distefano et al. 2002).   The estimated neutrino flux from GRBs implies that only nearby sources can be individually detected by the upcoming experiments.  However, the cumulative flux produced by the entire GRB population should be detectable
assuming that cosmological GEBs are the sources of the observed UHECRs (Waxman \& Bahcall, 1997).  This requires that the
rate of energy production of UHECRs in the energy interval $10^{19}-10^{21}$ eV is comparable to 
the rate of $\gamma$-ray production by GRBs in the BATSE band, $\dot{E}\sim10^{44.5}$ 
erg Mpc$^{-3}$ yr$^{-1}$ (Waxman 1995).


\acknowledgements This work was supported by an ISF grant for the Israeli Center for High Energy Astrophysics.


\end{document}